\documentclass[useAMS,usenatbib]{mn2e}

\usepackage{graphicx}

\title[]{The relationship between [OIII]$\lambda$5007\AA\ equivalent width 
and obscuration in AGN}
\author[A. Caccianiga and P. Severgnini]{A. Caccianiga$^{1}$\thanks{E-mail:
alessandro.caccianiga@brera.inaf.it} and P. Severgnini$^{1}$\\
$^{1}$INAF - Osservatorio Astronomico di Brera, via Brera 28, 
 I-20121 Milan, Italy}
\begin{document}

\date{}

\pagerange{\pageref{firstpage}--\pageref{lastpage}} \pubyear{2002}

\maketitle

\label{firstpage}

\begin{abstract}
In this paper we study the relationship between the equivalent width (EW) of 
the [OIII]$\lambda$5007\AA\ narrow emission line
in AGN and the level of obscuration. To this end, we combine the results of 
a systematic spectral analysis, both in the
optical and in the X-rays, on a statistically complete sample of $\sim$170
X-ray selected AGN from the XMM-{\it Newton} Bright Serendipitous Source sample
(XBS). 
We find that the observed large range of [OIII]$\lambda$5007\AA\ equivalent 
widths observed in the sample (from a few \AA\ up to 500\AA) is well explained 
as a combination of an intrinsic spread, probably due to the
large range of covering factors of the Narrow Line
Region, and the effect of absorption. The intrinsic 
spread is dominant for EW below 40-50\AA\ while absorption brings the values of
EW up to $\sim$100-150\AA\, for moderate levels of absorption (A$_V\sim$0.5-2 
mag) or up to $\sim$500\AA\ for A$_V>$2 mag. In this picture, 
the absorption has a significant impact on the observed EW also in type~1 AGN.
Using numerical simulations we find that this model is able to reproduce the 
[OIII]$\lambda$5007\AA\ EW distribution observed in the XBS sample and 
correctly predicts the shape of the EW distribution observed in the optically 
selected sample of QSO taken from the SDSS survey.
\end{abstract}


\section{Introduction}
The clouds of gas responsible for the narrow line emission
in Active galactic Nuclei (AGN), the so-called Narrow Line Region (NLR), 
are supposed to be placed at large distances from the central source, 
typically from tens of pc to kpc, where the potential well of the supermassive 
black-hole (SMBH) is too weak to induce high velocity motion. Nevertheless,
even at these large distances, the AGN emission is thought to be the main 
source of 
excitation for the  NLR clouds (e.g. see Osterbrock 1989). 
The NLR is often considered
as a good indicator of the isotropic emission of the AGN since it should 
reflect
the intensity of the nuclear source independently from its orientation in 
respect to 
the line-of-sight. Forbidden emission lines, in particular, like the 
[OIII]$\lambda$5007\AA\,  are very important
as they are not affected by a broad component produced in the Broad 
Line Region (BLR). 

It is well known (e.g. Baskin \& Laor 2005) 
that AGN show a large range of values (more than 2 orders of magnitude) 
of narrow line equivalent 
width (EW\footnote{We define equivalent width as: 
EW=$\int (F_\lambda /F_C-1)d\lambda$ where $F_C$ and $F_\lambda$ 
are the continuum and total flux respectively. 
Please note that the continuum flux includes both the AGN and the host-galaxy
flux.}).
This large spread has been usually interpreted in terms of 
different physical conditions of the NLR.
The NLR covering factor, in particular, has been considered the main
driver for the observed EW (Baskin \& Laor 2005). 
However, other factors can affect the 
EW of the narrow emission lines. For instance, it has been recently
proposed that the [OIII]$\lambda$5007\AA\ EW in type~1 AGN could also be 
sensitive to the orientation of the accretion disk 
(Risaliti, Salvati \& Marconi 2010).
Another potentially important effect that can influence the value of
EW is the absorption due to the molecular torus, postulated by the
unified model (Antonucci 1993). Indeed, since this kind of absorption
affects only the continuum emission and not the narrow emission line 
intensity, we expect a dependence of the observed EW with the amount
of obscuration. The fact that 
Seyfert~2 galaxies typically have [OIII]$\lambda$5007\AA\ with very 
large EW (up to 500\AA) supports this idea. 
Since moderate levels of absorption are
often observed also in type~1 AGN (e.g. De Zotti \& Gaskell 1985)
it is possible that this effect is
important for the global class of AGN and not just for 
the most absorbed ones.

In this paper, we want to quantify the importance of the absorption on 
the [OIII]$\lambda$5007\AA\ EW in all classes of AGNs, including type~1 
sources. 
To this end, we study the properties of the [OIII]$\lambda$5007\AA\ emission 
line  
in a well-defined flux-limited sample of X-ray selected AGN extracted from the
XMM-{\it Newton} Bright Serendipitous Sample (XBS, Della Ceca et al. 2004).
The peculiarity of this sample is that, thanks to its relatively bright
flux limit, the identification level is very high ($>$90\%) thus reducing any
possible bias related to the lack of some identifications. Furthermore, 
for all the AGN in the sample XMM-{\it Newton} X-ray data of medium/good
quality are available and a complete and systematic spectral analysis 
has been already carried out (Corral et al. 2011), 
providing important 
pieces of information like the intrinsic hydrogen column density (N$_H$) and 
the de-absorbed X-ray luminosity. We are thus in the position of combining
the [OIII]$\lambda$5007\AA\ properties (EW, luminosity) with a measure 
(or an upper limit) of the level of absorption, from the X-ray data. 

In Section~2 we describe the sample while in 
Section~3 we test whether the large spread of EW observed
in the AGN of the XBS sample can be due to the intrinsic
strenght of the NLR. In Section~4 
we study the impact of the obscuration on the observed 
[OIII]$\lambda$5007\AA\ EW. In Section~5 we
use numerical simulations to test in a quantitative way the hypothesis that
the absorption has an important role on the observed EW of the 
[OIII]$\lambda$5007\AA\ also in type~1 AGN. 
In Section~6 we compare the predictions of our
models with those of the disk-orientation model. The conclusions are 
summarized in Section~7.

Throughout the paper we assume $H_0$=65 km s$^{-1}$ Mpc$^{-1}$, 
$\Omega_{\lambda}$=0.7 and $\Omega_M$=0.3.

\section{The XBS sample}
The XMM-{\it Newton} Bright Serendipitous Survey (XBS survey, 
Della Ceca et al. 2004)
is a wide-angle ($\sim$28 sq. deg) high Galactic latitude ($|b|>$20 $\deg$) 
survey based on the XMM-{\it Newton} 
archival data. It is composed of two samples both flux-limited 
($\sim$7$\times$10$^{-14}$ erg cm$^{-2}$ s$^{-1}$) in two separate energy 
bands: the ``soft'' 0.5-4.5 keV band (the BSS sample) and the hard 4.5-7.5 keV 
band (the HBSS sample). A total of 237 (211 for the HBSS sample) 
independent fields have been used to select   400 sources, 
389 belonging to the BSS sample and 67 to the HBSS sample (56 sources are
in common). The details on the fields selection strategy, the source
selection criteria and the general properties of the 400 objects 
are discussed in Della Ceca et al. (2004). 

One of the main goals of the survey is to provide a well-defined and 
statistically complete census of the AGN population with particular 
attention to the problem of obscuration. To this end, the possibility 
of  comparing X-ray and optical spectra of good quality 
for all the sources present in the two complete samples offers 
a unique and fundamental tool to statistically study  the effect of absorption
in the AGN population in an unbiased way.
Indeed, most of the X-ray sources of the XBS survey 
have been detected with enough counts to allow a reliable X-ray  
spectral analysis. The systematic analysis of the X-ray spectra of all
the extragalactic sources is presented in Corral et al. (2011). 
At the same time, most of the sources 
have a relatively bright (R$<$22 mag) 
optical counterpart and they have been  
spectroscopically characterized using  4-meters-class telescopes.
To date,  the spectroscopic identification level has reached 
92\%. The results of the spectroscopic campaigns are discussed in 
Caccianiga et al. (2007, 2008). 

In total, the XBS sample contains about 300 AGN out of which 
169 have a redshift low enough (below $\sim$0.7) to allow the sampling
of the [OIII]$\lambda$5007\AA\ emission line in the optical spectrum.
This is the sample that we will analyse in the following sections.
In Sections 3 and 4 we will use AGNs from both BSS and HBSS sample
while, for the numerical simulations (Section~5) we will use only
the AGN in the BSS in order to have a simple and clear selection 
method that is an important requirement for this type of analysis.

\section{The effect of the intrinsic strength of the NLR}
Consistently with what is observed in other AGN samples, among the 
XBS AGN we find a very wide range of [OIII]$\lambda$5007\AA\ EW.
Considering all types of AGN the observed values of EW range from a few \AA\ up
to $\sim$500\AA. If we restrict the analysis just to the type~1 AGN
the range is somewhat smaller (from a few \AA\ up to $\sim$150\AA) but
still very broad. 
An intrinsic dispersion on the values of [OIII]$\lambda$5007\AA\ EW is
expected due to the variety of properties of the Narrow Line Region. 
For instance the NLR  covering factor 
is a parameter that is expected to influence the strenght of
the narrow emission lines in respect to the AGN continuum emission, and
therefore their equivalent widths (e.g. Osterbrock 1989; Baskin \& Laor 2005).

\begin{figure}
   \centering
    \includegraphics[width=9cm]{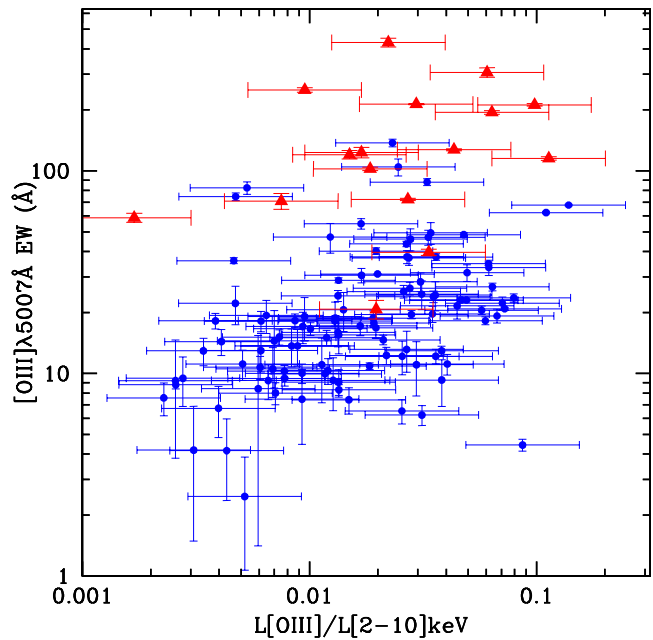}

   \caption{[OIII]$\lambda$5007\AA\ equivalent widths of the type~1 AGN (filled points -
blue in the electronic version)
and type~2 AGN (triangles - red in the electronic version) in 
the XBS versus the [OIII]$\lambda$5007\AA\ to X-rays
luminosity ratio.}
              \label{lumoiii_ew_ty1_ty2}
    \end{figure}

In order to test the hypothesis that the observed EW spread is related to
an intrinsically high luminosity of the NLR (in respect to the luminosity of
the central source) we have searched for correlation between the
[OIII]$\lambda$5007\AA\ EW of the AGN 
and the [OIII]$\lambda$5007\AA\ luminosity, 
normalized to the X-ray (de-absorbed) luminosity. The reason for 
using the de-absorbed X-ray luminosity as a reference is to have an 
independent indicator
of the AGN luminosity, already corrected for the absorption\footnote{
We note that, in case of large column densities, N$_H>$10$^{24}$ cm$^{-2}$, 
i.e. for the so-called Compton-thick AGNs, the computed 
X-ray luminosity could be strongly under estimated (of two 
orders of magnitude or more) since the absorption cut-off cannot be 
recognized in a 0.5-10 keV
spectrum. However, using the
available diagnostics we have shown in Corral et al. (2011) that no 
Compton-thick AGN candidates are present in the XBS sample.}.  
As shown in Fig.~\ref{lumoiii_ew_ty1_ty2} the [OIII]$\lambda$5007\AA\ EW in
type~1 AGN 
appears to be well correlated with the 
L[OIII]/L[2-10keV] ratio, at least up to EW$\sim$50-60\AA. This correlation
is confirmed by a Spearman's rank correlation analysis 
(see Tab.~\ref{correlation}).
In principle, the observed correlation could be - at least in part -
induced by the fact that both  [OIII]$\lambda$5007\AA\ EW and 
L[OIII]/L[2-10keV] ratio depend on the same quantity (L[OIII]). 
In order to test this possibilty
we have applied the partial correlation analysis (Kendall \& Stuart 1979)
that estimates the level of correlation between two variables excluding
the effect of a third variable. We find that the correlation between
[OIII]$\lambda$5007\AA\ EW and L[OIII]/L[2-10keV] is still present although
with a smaller significance (see Tab.~\ref{correlation})

On the contrary, if we consider only the 
type~2 AGN we do not observe any significant correlation.

%
\begin{table}

\caption{Values of probability for two uncorrelated variables to give
Spearman's rank correlation coefficient greater than the observed one.}
\label{correlation}      
\begin{tabular}{l r r r r}        
\hline\hline                 
                 & all      & Type 1  & Type 1   & Type 2 \\     
                 &          &         &  (low N$_H$)$^1$  &  \\
\hline           
EW vs L[OIII]/LX & $<$0.1\% & $<$0.1\%& $<$0.1\% & 25\%    \\
EW vs L[OIII]/LX &          &         &          &         \\
(excluding L[OIII])& 0.6\%     & 0.5\%    & 0.5\%     & 70\%    \\
EW vs N$_H$      & $<$0.1\% & 8\%  &  15\%      & 3\% \\
\hline                            
\end{tabular}
$^1$ N$_H<$10$^{21}$ cm$^{-2}$
\end{table}

These findings support the idea that the spread on the
[OIII]$\lambda$5007\AA\ EW in type~1 AGN is
mainly due to the NLR that can have a large range of luminosities
(up to 2 orders of magnitude) for a given AGN luminosity. The most 
obvious interpretation of this effect is that the covering factor 
of the NLR can vary significantly 
from source to source, as suggested by other authors 
(e.g. Baskin \& Laor 2005; Risaliti, Salvati \& Marconi 2010). 


%
At the same time, the fact that the correlation is not found for type~2
AGN indicates that other effects are regulating the values of EW. 
In the case of type~2 AGN the most obvious explanation is the 
absorption that affects the intensity of the continuum. 
The fact that also in type~1 AGN the correlation seems to
break down for EW larger than 40-50\AA\ suggests that a similar
mechanism could be at work also in this class of sources. In the next
section we analyse in detail this hypothesis. 

\section{The effect of absorption}
In this section we want to analyse the dependence of the 
[OIII]$\lambda$5007\AA\ 
EW with the level of absorption as derived from the X-rays. 
In Fig.~\ref{oiii_nh} we report the values of 
[OIII]$\lambda$5007\AA\ EW for all the AGN for which this
line is covered versus the value of N$_H$ computed through the
X-ray spectral analysis described in Corral et al. (2011). We have excluded 
from the plot the ``elusive'' AGN, i.e. the AGN whose spectrum is dominated
by the light from the host galaxy making their classification difficult or 
impossible (see Caccianiga et al. 2007 for details).  Even if with a 
great scatter, we observe a clear correlation (see Tab.~1)  
between the two parameters,
with the absorbed\footnote{
In Caccianiga et al
(2008) we have shown that the threshold A$_V$=2 mag, corresponding to 
N$_H$=4$\times$10$^{21}$ cm$^{-2}$ for a Galactic gas-to-dust ratio,  
well matches the optical separation between type~1 and type~2 AGN.} 
(N$_H>$4$\times$10$^{21}$ cm$^{-2}$) AGN showing the
largest values of [OIII]$\lambda$5007\AA\ EW (from 50 to 500\AA) and 
the unabsorbed/''weakly'' absorbed (N$_H<$4$\times$10$^{21}$ cm$^{-2}$) AGNs 
covering the
lowest values (between a few \AA\  up to $\sim$100\AA). The two groups
overlap mostly in the intermediate range of N$_H$ (10$^{21}$ cm$^{-2}$-
10$^{22}$ cm$^{-2}$). In agreement with what is expected from the unified
schemes, the correlation also holds using the optical classification 
(see Caccianiga et al. 2008 for details) instead of N$_H$, i.e.
the type~2 AGN show the largest values of [OIII]$\lambda$5007\AA\ EW while
type~1 AGN show the lowest values. The good agreement between optical and
X-ray ``classification'' observed in the XBS has been reported also previously 
(Corral et al. 2011; Caccianiga et al. 2004).
Notably, the correlation between [OIII]$\lambda$5007\AA\ EW and N$_H$ 
still holds (although marginally) when considering separately type~1 
and type~2 AGN while it disappears when we consider only the type~1 AGN
without  a significant absorption. This suggests that the effect of absorption 
could be relevant also within a single class of AGN.

The correlation observed between the [OIII]$\lambda$5007\AA\ EW and 
N$_H$/optical classification suggests a possible use of the 
[OIII]$\lambda$5007\AA\ EW as a simple proxy to classify a source as
type~1/type~2 or unabsorbed/absorbed AGN. This may be very useful 
for the identification of faint sources in deep surveys. 
For instance, by adopting a
limit on the EW of 100\AA\ type~1 and type~2 AGN, as well as 
unabsorbed/absorbed AGN, can be distinguished in our sample with a
good level of reliability ($>$70\%). The completeness, i.e. the
capability of including most of the sources of a given class, is
very high (80-90\%) for type~1 and unabsorbed AGN while it is around 60\%
for type~2 and absorbed AGN. Moreover, if we include in the computation 
also the ``elusive'' AGN, i.e. sources whose optical spectrum is dominated by
the host-galaxy, the capability of detecting
type~2/absorbed AGN just using the [OIII]$\lambda$5007\AA\ EW decreases 
dramatically
and the completeness falls down to $\sim$40\%. Therefore, caution must be
taken when using only the [OIII]$\lambda$5007\AA\ EW for the 
AGN classification.

\subsection{The model}
\begin{figure}
    \includegraphics[width=90mm]{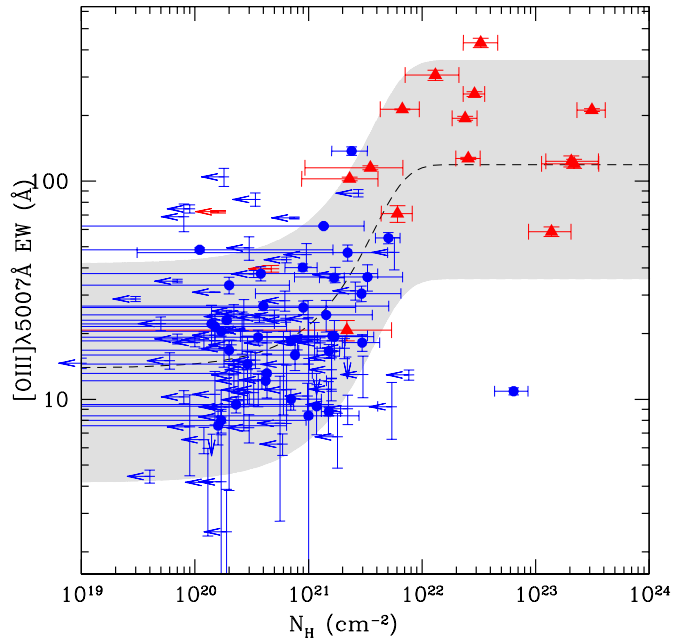}

   \caption{The [OIII]$\lambda$5007\AA\ equivalent width of the AGN in the
XBS sample versus the column densities derived from the X-ray data. Points 
(blue in the electronc version) and
triangles (red in the electronic version) indicate, respectively, 
the AGN optically classified as
type~1 and type~2. The dashed line represent the expected
values of EW as a function of N$_H$ according to the model described in the 
text 
assuming an AGN/galaxy luminosity ratio (at 4050\AA) of 15 and an AGN template 
with [OIII]$\lambda$5007\AA\ EW equal to 15\AA. The grey area 
represents  the expected value of EW when the starting value of 
[OIII]$\lambda$5007\AA\ EW is within a factor 3 in respect to the value in 
the adopted AGN template.}
              \label{oiii_nh}
    \end{figure}

In order to better understand the trend observed in Fig.~\ref{oiii_nh}, 
we have adopted a 
simple spectral model, described in Severgnini et al. (2003) and Caccianiga 
et al. (2008). This model uses an AGN template  composed of two parts: a) 
the continuum with the broad emission lines and b) the narrow emission lines.
According to the basic version of the AGN unified model, the first part can 
be absorbed while the second one is not affected by the presence of
an obscuring medium. The AGN template is based on the data taken 
from Francis et al. (1991) and Elvis et  al. (1994) while the extinction curve 
is taken from Cardelli, Clayton \& Mathis (1989). The EW of the 
[OIII]$\lambda$5007\AA\ 
emission line in the template is 15\AA. 
Besides the AGN  template, 
the spectral model includes also a galaxy template, produced on the basis of 
the Bruzual \& Charlot (2003) models. 
In order to set an average relative intensity of the AGN in respect to the 
host galaxy we have computed the average value of the absolute magnitudes 
observed for the type 1 AGN considered here 
($<M_R>\sim$-22.7) and assumed an average host-galaxy absolute magnitude
equal to M$^*_R$ (=$-$21, Brown et al. 2001). These two values yield an 
expected average AGN/galaxy ratio at 4050\AA\ (rest-frame) ($\eta_{4050}$) 
of $\sim$15. 
This parameter is expected to be very different, source by source, spanning
more than 2 orders of magnitude (see discussion in Caccianiga et al. 2007).
With the purpose of understanding the observed trend of Fig~\ref{oiii_nh} we
have fixed the AGN-to-host galaxy ratio to the expected average value. 
In principle, the value of $\eta$ in the average optical spectrum should be 
higher than
that estimated above because the spectrograph slit limits the amount of 
starlight from the host galaxy. However, at the average redshift of the
sources considered here ($<$z$>$$\sim$0.7) the slit width adopted for the
observations (typically 1.5$\arcsec$) correponds to $\sim$10 kpc and, 
therefore, most of the starlight is probably falling into the aperture making 
this correction, on average, negligible. 

\begin{figure}
   \centering
    \includegraphics[width=9cm]{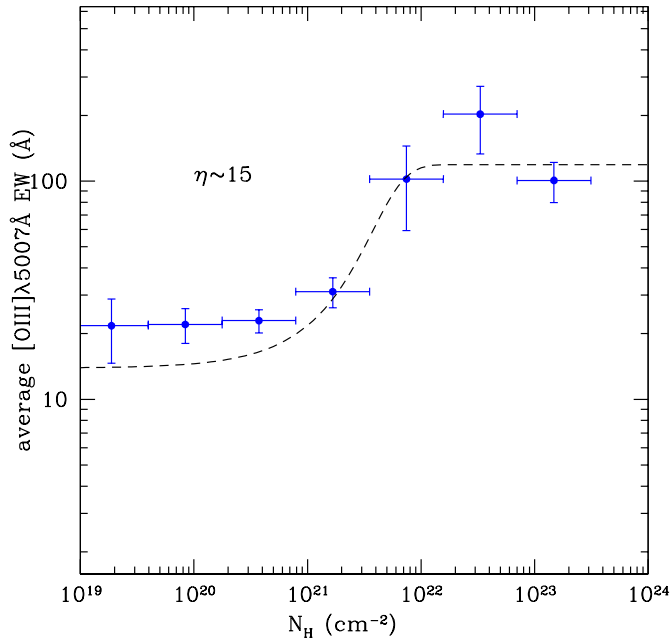}

  \caption{Distribution of the average values of [OIII]$\lambda$5007\AA\ 
for each bin of N$_H$. Errorbars indicate the errors on the average values.
In this plot we have considered the upper-limits on N$_H$ and on 
[OIII]$\lambda$5007\AA\ as detections. The dashed line is as in 
Fig~\ref{oiii_nh}.}
              \label{o_nh_binna}
    \end{figure}

We have then assumed different levels of N$_H$ and, adopting the Galactic
dust-to-gas ratio, we have derived the correspondent value of optical
absorption, A$_V$, and applied it to the template. 
In Fig.~\ref{oiii_nh} we plot the expected values of EW as a function of 
N$_H$. In the same figure
we have also plotted the regions where we expect to find sources if the
starting template has an intrinsic [OIII]$\lambda$5007\AA\ EW within a factor
3 in respect to the adopted AGN template. The width of this region is based on 
the intrinsic distribution derived in the previous section and represents
the expected range of EW before the effect of absorption.
For clarity we have also plotted the binned version of Fig.~\ref{oiii_nh} 
together with the model in a separate figure (Fig~\ref{o_nh_binna}).

Clearly, the
``S-shaped'' trend observed in the data points can be qualitatively reproduced
by the model: for low values of N$_H$ ($<$10$^{21}$ cm$^{-2}$) the 
[OIII]$\lambda$5007\AA\ EW is almost insensitive to the 
absorption and roughly equals the value adopted in the starting template;
as the value of absorption increases, the 
[OIII]$\lambda$5007\AA\ EW increases and ``saturates'' at high values 
(N$_H>$5$\times$10$^{22}$-10$^{23}$ cm$^{-2}$) due to the
presence of the emission from the host-galaxy which becomes important as
the absorbed AGN continuum gets weaker and weaker. In this case, the
value of EW converges to the limit value set by the adopted AGN-to-host galaxy
luminosity ratio and by the starting value of EW.


The evidences presented so far support a picture in which
the [OIII]$\lambda$5007\AA\ equivalent width is determined by a combination
of intrinsic intensity of the NLR in respect to the continuum 
(for EW$<$40-50\AA) and the effect of absorption (for EW$>$40-50\AA).
Notably, the effect of absorption is not important just for
type~2 AGN but it seems to have an important role also for type~1 AGN. 
However, due to the limited sensitivity of the X-ray data to column densities 
of
a few 10$^{21}$ cm$^{-2}$ or below (as shown by the range of upper limits in 
Fig.~\ref{oiii_nh}) 
it is difficult to test more quantitatively this picture through a direct
comparison of the values of N$_H$. Besides, it is reasonable to expect 
that the relationship between optical and X-ray absorption has an 
intrinsic dispersion (see, for instance, Maiolino et al. 2001 and  Maiolino, 
Marconi, 
\& Oliva 2001) and, as a consequence, that the value of N$_H$
can be used as an indicator of the optical absorption only in a 
statistical sense. 
If we split the sample of type~1 AGN in two groups with 
[OIII]$\lambda$5007\AA\ EW larger or lower than 40\AA\ and compute the
average values of N$_H$ (considering just the detections and excluding the
source with very high column densities, $>$5$\times$10$^{22}$ cm$^{-2}$)
we obtain a marginal ($\sim$1.4$\sigma$) evidence that the 
type~1 AGN with large EW have also larger 
($<N_H>=$2.0$\pm$0.8$\times$10$^{21}$ cm$^{-2}$) column densities when compared
to low-EW type~1 AGN ($<N_H>=$0.9$\pm$0.2$\times$10$^{21}$ cm$^{-2}$).

In order to test more quantitatively this picture we have used a simple
numerical simulation that we describe in the following section.

\section{Numerical simulations}

The fact of that we measure the level of absorption from the X-rays and that
there is an intrinsic dispersion between optical and X-ray absorption, prevents
us to directly test, source by source, the idea that the absorption is 
relevant also in type~1 AGNs. We thus decide to approach the problem in a 
statistical way. If A$_V$ and N$_H$ are linearly correlated (although with an 
intrinsic dispersion) we expect that the A$_V$ distribution statistically 
follows the distribution of N$_H$. Therefore, even if we are not able to 
determine univocally the value of A$_V$ of a single source, we can reasonably
assume to know the distribution of A$_V$ if we know the distribution of N$_H$.
Using this distribution, we now want to assess whether we are able to reproduce
the observed distribution of EW. 


 \begin{figure}
   \centering
    \includegraphics[width=9cm]{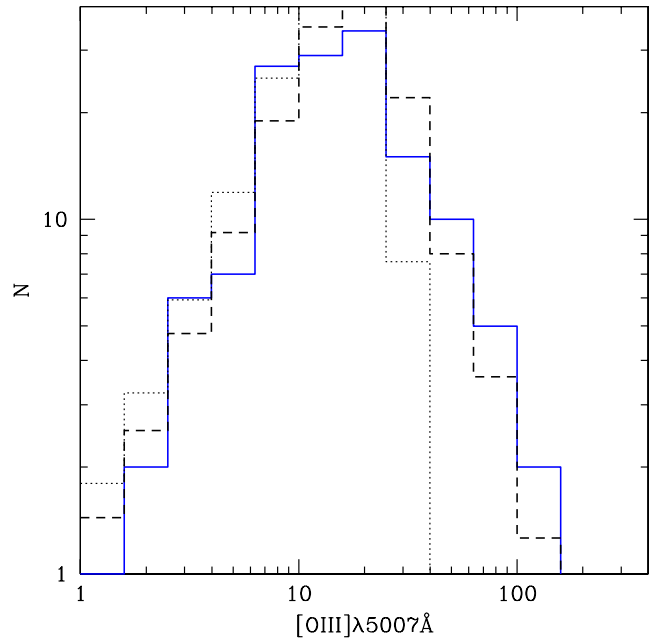}

   \caption{Comparison between the distribution of the 
[OIII]$\lambda$5007\AA\ equivalent widths of the BSS (continuous line) 
and the simulated sample (dashed line) of Type~1 AGNs. 
The intrinsic distribution
of EW (i.e. before the absorption) is indicated with the dotted line.}
              \label{simul_obs}
    \end{figure}

As starting point we need a form for the intrinsic dispersion that, as 
explained in the previous sections, is likely to dominate the low
tail of the EW distribution ($<$40\AA). To this end we have fitted the 
low tail of the observed EW distribution with a Gaussian profile 
peaked around EW=12\AA\ 
(a value very close to the one measured in the AGN template by 
Francis et al. 1991) and
derived a sigma of $\sim$8\AA.

We then use the distribution of
N$_H$ observed in type~1 AGNs 
of the BSS sample (i.e. the one selected
in the 0.5-4.5 keV band).  
Since, as said before, we have many upper limits on N$_H$ in our sample
we have used the method described in Avni et al. (1980) 
to derive the ``real'' N$_H$ distribution.
We have found that the distribution (in Log N$_H$) is best represented 
by two Gaussian, one peaked at 
Log N$_H$=20.15 cm$^{-2}$ and $\sigma$=0.5  
and one peaked at Log N$_H$=21.2 cm$^{-2}$ and $\sigma$=0.35. 
The second Gaussian has a relative normalization of 0.44 times in respect the
first one. As explained above, we assume that this distribution 
coincides with the distribution of A$_V$, once convereted using 
the Galactic dust-to-gas ratio. 
We then extract randomly one EW from the intrinsic distribution and one
value of A$_V$ from the A$_V$ distribution and ``absorb'' the continuum
below the line accordingly.
Following our classification criteria, in the simulation we consider an AGN 
as type~1 only if the value of Log N$_H$ is below 21.6. 

In Fig.~\ref{simul_obs} we compare the simulated EW distribution for type~1 
AGN with the one observed in the BSS sample. 
The good agreement is confirmed by an independent K-S test
(probability $>$10\%).

It could be questioned that the (low) absorptions observed in 
type~1 AGN may be related (at least in part) to the host-galaxy and not
to the molecular torus. This is a concrete possibility for values 
below 10$^{21}$ cm$^{-2}$. 
In this case, the effect of the absorption on the [OIII]$\lambda$5007\AA\
EW is less obvious. If the dust from the host galaxy influences equally 
both the continuum and the NLR
emission then we expect that the net effect on the equivalent width of 
[OIII]$\lambda$5007\AA\ is null. Under this condition the model described 
here, that assumes a 
different effect of the absorption on the continuum and the emission line 
flux, cannot be applied. If, on the contrary, the host galaxy is not able to
completely hide the NLR, as suggested by some studies (e.g. Keel 1980; 
De Zotti and Gaskell 1985), then we must expect a net effect on the  
[OIII]$\lambda$5007\AA\ EW similar to that predicted from the dusty torus.

As described above, the observed distribution of N$_H$ 
in the type~1 AGN of the BSS sample is well described by two gaussians centered
respectively to Log N$_H$=20.15 and 21.2 cm$^{-2}$ and a relative 
normalization of $\sim$2:1. It is therefore possible that the first Gaussian
is (mainly) due to the host-galaxy and that only the second  one is 
produced by the 
torus. We have thus modified the simulations by assuming that only the
second Gaussian has the effect on the [OIII]$\lambda$5007\AA\ EW. 
We find that the results of the simulation do not change since only the
high part of the N$_H$ distribution ($>$10$^{21}$ cm$^{-2}$) has a relevant
effect on the distribution of [OIII]$\lambda$5007\AA\ EW.

We conclude that the observed high values of [OIII]$\lambda$5007\AA\ EW
($>$40\AA) in type~1 AGN can be consistently 
explained as the result of the effect
of a mild obscuration 
(10$^{21}$ cm$^{-2}<$N$_H<$4$\times$10$^{21}$ cm$^{-2}$) probably due to  
the molecular torus.

\section{The effect of disk-orientation}

In a recent paper, Risaliti, Salvati \& Marconi (2010) have suggested that
the distribution of [OIII]$\lambda$5007\AA\ EW in type~1 AGN contains 
information on the orientation of the accretion disk in respect to the
line-of-sight. Indeed, since the emission from a disk is not isotropic, the
observed optical continuum is 
expected to decrease for large observing angles (in respect to the normal)
while the line flux is supposed not to vary with the inclination. As
a consequence, we expect a variation of the [OIII]$\lambda$5007\AA\ EW 
with the accretion disk orientation. 
By means of a comparison between the [OIII]$\lambda$5007\AA\ EW observed in a 
sample of $\sim$6000 type~1 AGN of the SDSS and the results of numerical 
simulation, these authors have concluded that the values of EW between 40\AA\ 
and $\sim$100\AA\ are likely the consequence of disk-inclination effect 
while the values below 40\AA\ are probably due to the
intrinsic scatter related to the NLR geometry/covering factor. 
This result, if confirmed, would imply that the orientations of
the molecular torus and of the disk are not coupled or that the 
torus opening angle is very large otherwise it would not
be possible to observe high disk inclinations because of the presence of
the obscuring medium. 
According to Risaliti, Salvati \& Marconi (2010), the strong evidence for
the disk-inclination model is the slope (=$-$3.5) observed in the 
EW distribution above 40-50\AA. 
This characteristic slope can be well reproduced
by their numerical simulation if the effect of the presence of a 
flux/luminosity limit is taken into account.


 \begin{figure}
   \centering
    \includegraphics[width=9cm]{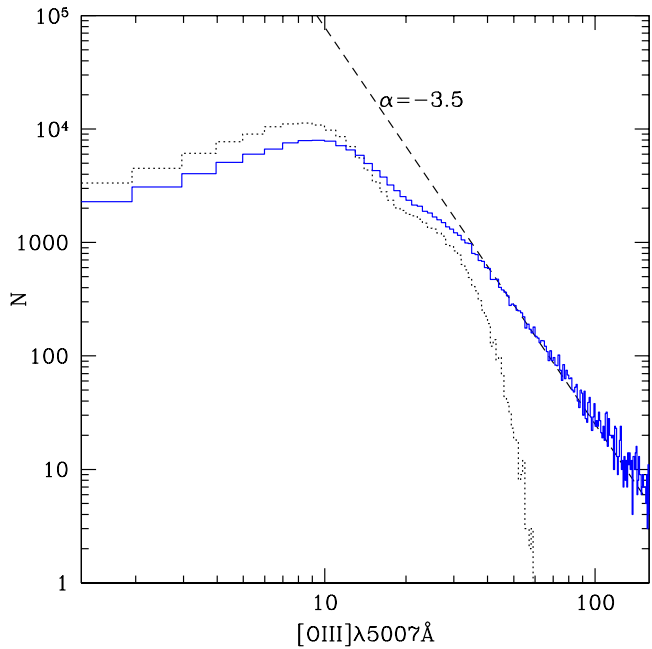}

   \caption{The simulated [OIII]$\lambda$5007\AA\ EW distribution 
for type~1 AGN using the shape of the intrinsic distribution found by
Risaliti, Salvati \& Marconi (2010) for the SDSS type~1 AGN sample. 
The dashed line represents the slope 
of the tail at high EW observed in the SDSS sample of type~1 AGN discussed
in Risaliti, Salvati \& Marconi (2010). Please note that, in order to 
facilitate the comparison with the plots in Risaliti, Salvati \& Marconi 
(2010) the binning adopted in this figure is linear while in the previous 
figure the binning is logarithmic.}
              \label{simula_ty1_risaliti}
    \end{figure}
 
In the previous sections we have presented evidences that the 
``tail'' above 40\AA\ of the [OIII]$\lambda$5007\AA\ EW distribution in the
type~1 AGN of the BSS sample could be due to the effect of the absorption 
(likely related to the molecular torus). It is thus interesting to evaluate if this 
effect can be important also for the SDSS sample. 
As explained in the previous section, the critical values of N$_H$ 
are those above 10$^{21}$ cm$^{-2}$. 
In principle it is not clear whether the selection criteria adopted in 
the SDSS sample have an impact on the N$_H$ distribution, for type~1 AGN. 
Therefore, 
it is important first of all to check whether in the SDSS sample used by
Risaliti, Salvati \& Marconi (2010) there are sources with N$_H>$10$^{21}$ 
cm$^{-2}$. 
In Young, Elvis \& Risaliti (2009) it is presented the X-ray analysis
of a subsample of the SDSS quasar sample containing the sources falling
serendipitously in XMM-{\it Newton} fields. Considering only the 
$\sim$120 sources with redshift below 0.8 (to match the sample
used by Risaliti, Salvati \& Marconi 2010) we count 10 sources 
with a detected value of N$_H>$10$^{21}$ cm$^{-2}$ plus 26 sources
with an upper limit on N$_H$ above 10$^{21}$ cm$^{-2}$. Therefore, the 
percentage of AGNs in the SDSS with N$_H>$10$^{21}$ cm$^{-2}$ 
ranges from 8\% up to 30\%. In the BSS sample we have estimated that 
the fraction of type~1 AGN with N$_H>$10$^{21}$ cm$^{-2}$ is similar 
(about 20\%,
taking into account the upper limits through the survival analysis).
We thus expect that also in the SDSS sample the impact of the absorption on the
[OIII]$\lambda$5007\AA\ EW distribution is important. 
We now want to quantify this statement by using the numerical simulations 
described in the previous section. In particular, we want to see whether
the slope of -3.5 observed in the SDSS sample can be similarly reproduced also
by a model where the absorption, and not the disk-orientation, is the
principal driver for the observed EW distribution. 

The intrinsic distribution, important for both models below 40\AA, has
a different shape/parametrization in the BSS and the SDSS sample. As
explained above we have used a single Gaussian centered at 12 \AA\ and
a $\sigma$=8\AA\, while Risaliti, Salvati \& Marconi (2010) used two
Gaussians. 
For consistency with the work by Risaliti, Salvati \& Marconi (2010) 
we have adopted the same intrinsic distribution used in their paper.
The second piece of information required  
to run the simulations is the observed distribution of N$_H$. 
Since the fraction of type~1 AGN with N$_H>$10$^{21}$ cm$^{-2}$ is
similar in the two sample it is reasonable to use the same N$_H$ distribution 
derived from the BSS sample.

In Figure~\ref{simula_ty1_risaliti} we have reported the results of the
simulations. As a reference, it is also plotted the slope -3.5 observed
in the SDSS sample. It is clear that: 1) the presence of sources with
large values of N$_H$ is expected to produce a significant tail of
EW above 40-50\AA\ and 2) the predicted slope for EW$>$40-50\AA\ is
consistent with the one observed in the SDSS sample. 
We conclude that the hypothesis that the absorption is the main 
driver for the observed EW of the [OIII]$\lambda$5007\AA\ could be valid 
also for the SDSS sample of type~1 AGN. It is also possible that both
absorption and disk-orientation effects co-exist in the same sample. 
In this case, in order to infer information on the disk inclination it is
necessary to properly take into account the effect of absorption on the
[OIII]$\lambda$5007\AA\ EW.

\section{Summary and conclusions}

We have studied the equivalent widths of the 
[OIII]$\lambda$5007\AA\ narrow emission line in a sample of $\sim$170 X-ray 
selected AGN taken from the XMM-{\it Newton} Bright Survey. 
We have combined optical and X-ray information in order to understand
the origin of the observed large range of [OIII]$\lambda$5007\AA\ EW values 
(from a few \AA\ up to 500\AA) and, in particular, to quantify the
importance of absorption. 
The main results can be summarized as follows:

\begin{itemize}

\item The values of EW correlate with the luminosity of the 
[OIII]$\lambda$5007\AA\ normalized to the (de-absorbed) X-ray luminosity, at 
least up to EW$\sim$40-50\AA. 
This suggests that the values of [OIII]$\lambda$5007\AA\ from $\sim$1\AA\ up to
40-50\AA\ are determined by
the intrinsic strength of the NLR (related, for instance, to the NLR covering
factor). Above EW=40-50\AA\ the correlation does not hold anymore. In this 
range of EW we have both type~1 and type~2 AGNs;

\item We have found a strong dependence of the [OIII]$\lambda$5007\AA\ EW
with the absorbing column density, N$_H$, derived from the X-ray spectra. 
On average, sources with low values of N$_H$ ($<$10$^{21}$ cm$^{-2}$) 
present low average values of [OIII]$\lambda$5007\AA\ EW ($\sim$20\AA) 
while 
sources presenting large column densities (N$_H>$10$^{22}$ cm$^{-2}$) 
have the largest values of [OIII]$\lambda$5007\AA\ EW  
(up to $\sim$500\AA).
This trend is well explained in the context of the AGN unified model,
which postulates a connection between optical and X-ray absorption
(due to the putative molecular torus): in this framework the increase
of the [OIII]$\lambda$5007\AA\ EW with N$_H$ is simply caused by the decrease
of the optical continuum beneath the line due to its progressive 
obscuration. In this sense, the value of [OIII]$\lambda$5007\AA\ EW is
a rough indicator of the orientation of the AGN (in respect to the
torus opening angle) and can be used as simple proxy for the 
optical/X-ray classification of an AGN (type1/type2 or unobscured/obscured
AGN) although not very efficient in identifying all the type~2/absorbed
AGNs (completeness level $\sim$40\%);

\item By means of numerical simulations we have demonstrated that the
observed distribution of [OIII]$\lambda$5007\AA\ EW of type~1 AGN 
is well reproduced by
combining an intrinsic distribution (a symmetric Gaussian centered at
EW=12\AA\ and a $\sigma$=8\AA) with the distribution of absorption 
derived from the X-ray analysis. Therefore, we conclude that 
also in type~1 AGN the effect of absorption is important and 
produces [OIII]$\lambda$5007\AA\ lines with large EW (between 40\AA\ and 
100\AA).

\item The combination of intrinsic EW distribution and mild (N$_H$ of a few 
10$^{21}$ cm$^{-2}$) absorption 
seems to be a valid explanation 
also for what is observed in the SDSS sample considered by Risaliti, Salvati \&
Marconi (2010) and thus represents an alternative to the
disk-inclination hypothesis, although we cannot exclude that both effects
are at work in the sample.

\end{itemize}

\section*{Acknowledgments}
We thanks Marco Salvati and Guido Risaliti for useful discussions. 
We also thanks the referee, Robert Antonucci, for his comments that
improved the manuscript. 
The authors acknowledge financial support from ASI
(COFIS contract and ASI-INAF grant n. I/009/10/0).
{}

\end{document}